\documentclass[aps,pra]{revtex4}
\usepackage{graphicx}

\begin{document}

\title{Classical communication and entanglement cost in preparing a class of multi-qubit states}
\author{Zhan-jun Zhang$^{a,c,*}$, Yi-min Liu$^b$, Dong Wang$^a$\\
{\normalsize $^a$ Key Laboratory of Optoelectronic Information
Acquisition \& Manipulation of Ministry of Education of China,}\\
{\normalsize School of Physics \& Material Science, Anhui University, Hefei 230039, China} \\
{\normalsize $^b$ Department of Physics, Shaoguan University, Shaoguan 512005, China }\\
{\normalsize $^c$ Department of Physics and Center for Quantum Information Science,}\\
{\normalsize National Cheng Kung University, Tainan 70101, Taiwan}\\
{\normalsize $^{*}$ Corresponding author's Email: zjzhang@ahu.edu.cn
(Z. J. Zhang)}}

\maketitle

\begin{minipage}{430pt}

{\bf Abstract} Recently, several similar protocols[J. Opt. B 4
(2002) 380; Phys. Lett. A 316 (2003) 159; Phys. Lett. A 355 (2006)
285; Phys. Lett. A 336 (2005) 317] for remotely preparing a class
of multi-qubit states (i.e, $\alpha|0 \cdots
0\rangle+\beta|1\cdots 1\rangle$) are proposed, respectively. In
this paper, by applying the controlled-not (CNOT) gate, a new
simple protocol is proposed for remotely preparing such class of
states. Compared to the previous protocols, both classical
communication cost and required quantum entanglement in our
protocol are remarkably reduced. Moreover, the difficulty of
identifying some quantum states in our protocol is also degraded.
Hence our protocol is more economical and feasible. \\


\end{minipage}

Quantum entanglement and classical communication are two
elementary resources in quantum information field. With these two
elementary resources and certain local unitary operations, lots of
interesting and important works have been done[1-8], such as
quantum teleportation[1], quantum secret sharing (QSS)[2], quantum
key distribution(QKD)[3], etc. In 2000, a distinct application of
quantum entanglement and classical communication, i.e., remote
state preparation (RSP), was proposed by Lo[9]. In RSP by means of
a prior shared entanglement and some classical communication, a
pure known quantum state is prepared in a remote place via certain
local unitary operations. Different from quantum teleportation, in
RSP the sender Alice is assumed to completely know the state to be
prepared remotely. Due to the prior knowledge about the quantum
state, to some extend the classical communication and entanglement
cost can be reduced in RSP process. For an example, Pati [10] has
shown that for a qubit chosen from equatorial or polar great
circles on a Bloch sphere, RSP requires only 1 forward classical
bit from Alice to Bob, exactly half that of teleportation. On the
other hand, RSP exhibits a stronger trade-off relation between the
required entanglement and the classical communication cost than
quantum teleportation. The RSP protocols proposed by Lo[9] and
Bennett et al.[11] show that in the presence of a large amount of
previously shared entanglement, the asymptotic classical
communication cost of RSP for general states is one bit per qubit.
However, for special states, RSP protocol is more economical than
quantum teleportation.

So far, various RSP protocols have been put forward in many
literatures[12-33]. Among these protocols, some[29-34] concentrate
on the preparation of a class of multi-qubit state (i.e.,
$\alpha|0 \cdots 0\rangle+\beta|1\cdots 1\rangle$). Specifically,
in 2002, Shi et al.[29] proposed a scheme for remotely preparing a
two-qubit entangled state (i.e.,
$\alpha|00\rangle+\beta|11\rangle$) by consuming a three-qubit
Greenberger-Horne-Zeilinger (GHZ) state and 1-bit classical
communication; In 2003, by means of two pairs of entangled qubits
and 2 classical bits, Liu et al.[30] proposed other scheme for
preparing remotely a two-qubit entangled state (i.e.,
$\alpha|00\rangle+\beta|11\rangle$); Very recently, Dai et al.[31]
proposed a RSP scheme of a four-qubit GHZ class state (i.e.,
$\alpha|0000\rangle+\beta|1111\rangle$) via two non-maximally
entangled three-qubit GHZ states, and the necessary classical
communication cost is 1 bit; Wang et al.[33] proposed a RSP
protocol of a three-qubit state (i.e.,
$\alpha|000\rangle+\beta|111\rangle$) by taking both a three-qubit
entangled state and a two-qubit entangled state as the quantum
channel, and the necessary classical communication cost is 0.5
bits on average. Moreover, the clone protocol proposed by Zhan[34]
is virtually a RSP protocol of a two-qubit entangled state (i.e.,
$\alpha|00\rangle+\beta|11\rangle$) with two Bell states and 2
classical bits. In this paper, we will propose a new simple
protocol for remotely preparing a class of multi-qubit states
(i.e., $\alpha|0 \cdots 0\rangle+\beta|1\cdots 1\rangle$).
Compared to these previous protocols[29-31, 33-34], both required
quantum entanglement and classical communication cost in our this
protocol are greatly reduced, as means that our protocol is more
economical. This is a distinct advantage of our protocol and one
will see it later.

For simplicity, let us firstly consider the remote  preparation of a
two-qubit entangled state (i.e., $\alpha|00\rangle+\beta|11\rangle$)
with a Bell state as the quantum channel. Suppose Alice is the
ministrant while Bob the state preparer. Alice wants to help Bob to
prepare remotely a two-qubit entangled state
$|P\rangle=\alpha|00\rangle+\beta|11\rangle$, where $\alpha$ is real
and $\beta$ is complex and $|\alpha|^{2}+|\beta|^{2}=1$. Bob dose
not know the two coefficients of the state $|P\rangle$ but Alice
dose. In addition, as mentioned before, the state taken as the
quantum channel between Alice and Bob is a Bell state. Without loss
of generality, we take it as
\begin{eqnarray}
|\phi\rangle_{12}=\frac{1}{\sqrt{2}} (|00\rangle+|11\rangle)_{12}.
\end{eqnarray}
Suppose qubit 1 belongs to Alice while qubit 2 to Bob. In order to
help Bob to prepare the state $|P\rangle$, Alice performs a
single-qubit projective measurement on her qubit 1 in a set of two
mutually orthogonal basis vectors $\{|\psi\rangle,
|\psi_{\perp}\rangle\}$, where
\begin{eqnarray}
|\psi\rangle=\alpha|0\rangle+\beta|1\rangle,\ \ \ \
|\psi_{\perp}\rangle=\beta^{*}|0\rangle-\alpha|1\rangle.
\end{eqnarray}
Note that $|\psi\rangle$ is exactly the original state
$|P\rangle$. These two mutually orthogonal basis vectors are
related to the computation basis vectors $\{|0\rangle,
|1\rangle\}$ and form a complete orthogonal basis set in a
sing-qubit 2-dimensional Hilbert space. Then we observe the shared
Bell state between Alice and Bob can be rewritten as
\begin{eqnarray}
|\phi\rangle_{12}&=&\frac{1}{\sqrt{2}}[|\psi_{\perp}\rangle_1(\beta|0\rangle-\alpha|1\rangle)_2
+|\psi\rangle_1(\alpha|0\rangle+\beta^*|1\rangle)_2].
\end{eqnarray}
According to the equation (3), one can see that Alice's
single-qubit projective measurement result should be
$|\psi_{\perp}\rangle_1$ or $|\psi\rangle_1$. For each measurement
result, its occurrence probability is $1/2$. Incidentally, before
Alice informs of Bob her measurement result, they agree that
$|\psi_{\perp}\rangle_1$ corresponds to the classical bit "0".

If the measurement result is $|\psi_{\perp}\rangle_1$, Alice sends
a classical bit "0" to Bob. After having received Alice's
classical bit "0" in a certain interval, Bob knows that Alice's
measurement result is $|\psi_{\perp}\rangle_1$ and the state of
qubit 2 has collapsed to $(\beta|0\rangle-\alpha|1\rangle)_2$.
Then to realize the preparation of the two-qubit entangled state
$|P\rangle$, Bob carries out some actions as following.

{\it Step (1)} Bob performs the unitary operation
$U_1=(|1\rangle\langle0|-|0\rangle\langle1|)_2$ on his qubit 2.
After his this unitary operation, the collapsed state of qubit 2
$(\beta|0\rangle-\alpha|1\rangle)_2$ is transformed into
\begin{eqnarray}
|T\rangle_2=(\alpha|0\rangle+\beta|1\rangle)_2.
\end{eqnarray}

{\it Step (2)} Bob introduces an auxiliary qubit 3 in the state
$|0\rangle$. He performs a controlled-not (CNOT) gate operation
$C_{2,3}$ by taking the qubit 2 as a control qubit and the qubit 3
as a target one. This unitary operation transforms the joint state
of the qubits 2 and 3 as following
\begin{eqnarray}
C_{2,3}|T\rangle_2|0\rangle_3=(\alpha|00\rangle+\beta|11\rangle)_{23}.
\end{eqnarray}
One can see that after the unitary operation the original state
$|P\rangle$ has been successfully prepared in the qubits 2 and 3.

If Alice's measurement result is $|\psi\rangle_1$, the state of
qubit 2 collapses to $(\alpha|0\rangle+\beta^*|1\rangle)_2$. Since
Bob has no knowledge about the coefficients $\alpha$ and $\beta$,
it seems that he can not convert this collapsed state to the state
$\alpha|0\rangle+\beta|1\rangle$ via a certain unitary operation.
This means Bob can not prepare the original state $|P\rangle$ by
applying the CNOT gate. Apparently, Alice needs not to send any
classical bit to Bob. However, as mentioned before, the
coefficient $\alpha$ is assumed real while $\beta$ complex in the
beginning. Then one would like to ask whether the conversion can
be realized in the case of the latter outcome provided that Alice
lets Bob know some information about the coefficients. In fact,
there really exist two special $cases$: $(A)$  Both $\alpha$ and
$\beta$ are real; $(B)$ $\alpha$ is $\frac{1}{\sqrt{2}}$ and
$\beta$ is $\frac{1}{\sqrt{2}}e^{i\theta}$ ($\theta$ is an
arbitrary real parameter). In these two special cases, for the
latter outcome, the preparation of the original state can also be
realized at Bob's place.  This can be seen from the following
detailed analysis process. Incidentally, in the two special cases,
two classical bits "10" or "11" instead of the single classical
bit "0" are sent from Alice to Bob to realize his preparation. The
first classical bit "1" corresponds to Alice's measurement result
$|\psi\rangle_1$ and the second bit "0" ("1") to $Case (A)$
($(B)$).

$Case (A)$ Both $\alpha$ and $\beta$ are real. After obtaining the
measurement result $|\psi\rangle_1$, Alice sends Bob two classical
bits "10" in a certain interval. According to these information,
Bob knows that Alice's measurement result is $|\psi\rangle_1$ and
both $\alpha$ and $\beta$ are real. Then he can infer that the
state of qubit 2 has collapsed to
$(\alpha|0\rangle+\beta|1\rangle)_2$. In order to achieve his goal
of preparing remotely the state $|P\rangle$, Bob repeats the step
(2) proposed above. After this, Bob has successfully realized the
preparation of the state $|P\rangle$ at his place. Therefore, in
the case that both $\alpha$ and $\beta$ are real, the original
state $|P\rangle$ can always be deterministically prepared in the
remote place.

$Case (B)$ $\alpha$ is $\frac{1}{\sqrt{2}}$ and $\beta$ is
$\frac{1}{\sqrt{2}}e^{i\theta}$ ($\theta$ is an arbitrary real
parameter). If Alice gets $|\psi\rangle_1$, she sends two
classical bits "11" to Bob. According to this information, Bob
knows that the state of qubit 2 has collapsed to
$\frac{1}{\sqrt{2}}(|0\rangle+e^{-i\theta}|1\rangle)_2$. To
prepare the original state $|P\rangle$ in his place, Bob firstly
performs the unitary operation
$U_2=(|0\rangle\langle1|+|1\rangle\langle0|)_2$, which transforms
the above collapsed state into
$e^{-i\theta}\times\frac{1}{\sqrt{2}}(|0\rangle+e^{i\theta}|1\rangle)_2$,
and then he repeats the step (2). After this, the joint state of
qubits 2 and 3 in his possession is
\begin{eqnarray}
e^{-i\theta}\times\frac{1}{\sqrt{2}}(|00\rangle+e^{i\theta}|11\rangle)_{23}.
\end{eqnarray}
It is exactly the original state $|P\rangle$ except for an overall
unimportant phase factor. This means for the special values of
$\alpha$ and $\beta$, the remote preparation can also be
deterministically realized.

By the above analyses, one can easily conclude that with one Bell
state and one auxiliary qubit, the original state $|P\rangle$ can be
prepared probabilistically or deterministically in a remote place.
In a general case, the total success probability is $\frac{1}{2}$
(probabilistic) and the necessary classical communication cost is
0.5 ($\frac{1}{2}\times 1 + \frac{1}{2}\times 0=0.5$) forward bit
{\bf on average}. Nonetheless, if the state $|P\rangle$ belongs to
the special states described in cases $(A)$ and $(B)$, the success
probability of preparation can achieve 1 (deterministic), and the
classical communication cost is increased to 1.5 ($\frac{1}{2}\times
1+\frac{1}{2}\times 2=1.5$) bits {\bf on average}. {\it It should be
emphasized that the fraction of bit as the classical communication
cost is not created first by this paper but has already occurred in
Ref.[31].}

Now let us generalize the above protocol to a multi-qubit state
case, i.e., the state to be prepared remotely is a $m(
m\geq3)$-qubit state
\begin{eqnarray}
|P'\rangle=\alpha\prod^m_{i=1}|0\rangle_i+\beta\prod^m_{i=1}|1\rangle_i.
\end{eqnarray}
In this case, the demonstration of preparing the $m$-qubit state
is very similar to the above process except for a little
modification in the step (2). That is, after having got the
collapsed state $(\alpha|0\rangle+\beta|1\rangle)_2$, instead of
introducing one auxiliary qubit, Bob introduces $m$-1 auxiliary
qubits each in the state $|0\rangle$. To achieve his goal of
preparation, Bob performs the CNOT gate operations
$C_{2,m+1}\cdots C_{2,4}C_{2,3}$ with the qubit 2 always as the
control qubit and the other qubits as the target qubits. In this
way, after having performed the local CNOT gate operations, the
joint state
$(\alpha|0\rangle+\beta|1\rangle)_2\prod^{m+1}_{k=3}|0\rangle_k$
of the qubits of 2, 3, ..., $m$+1 becomes
\begin{eqnarray}
&&C_{2,m+1}\cdots C_{2,4}C_{2,3}
(\alpha|0\rangle+\beta|1\rangle)_2\prod^{m+1}_{k=3}|0\rangle_k\nonumber\\
&=&(\alpha|00...0\rangle+\beta|11...1\rangle)_{23...m+1}=\alpha\prod^{m+1}_{k=2}|0\rangle_k+\beta\prod^{m+1}_{k=2}|1\rangle_k.
\end{eqnarray}
It is exactly the original state $|P'\rangle$ which needs to be
prepared remotely. This means Bob has already successfully
prepared the state $|P'\rangle$ at his place.

Similarly, in the process of preparing a multi-qubit state, Alice
will get the measurement result $|\psi\rangle_1$ with probability
$\frac{1}{2}$. In this case, due to the same reason mentioned
previously, it seems that the original state $|P'\rangle$ cannot
be prepared remotely and Alice needs not to send any classical bit
to Bob. Likewise, there also exist two exceptions, i.e., the two
coefficients of the state $|P'\rangle$ belongs to some special
values as described in the $cases (A)$ and $(B)$. In these two
cases, applying the same method proposed previously, the remote
preparation of the state $|P'\rangle$ can also be
deterministically realized by applying the CNOT gate $m$-1 times.

Based on above analyses, one can easily see that, even in the case
that the state to be prepared is a $m$-qubit ($m\geq3$) state, by
means of one Bell state and 0.5 classical bits on average, the
preparation of the original state can also be realized
probabilistically, and the success probability is at least
$\frac{1}{2}$. The additional cost is that $m$-1 auxiliary qubits
should be introduced and the local unitary operation of CNOT gate
should be applied $m$-1 times. Similarly, if the state
$|P'\rangle$ belongs to some special states, the remote
preparation of the original state can be realized in a
deterministic manner. The necessary classical communication cost
is increased to 1.5 bits on average. Incidentally, it is obvious
that the present RSP protocol of a class of multi-qubit state can
be simply and straightforwardly generalized to the case that
partial entangled state instead of Bell state is taken as quantum
channel. \\

Table I: The comparisons between our protocol and the previous
protocols. Q.S. denotes the quantum state to be prepared remotely
in the protocol; S.Q.C. denotes the states taken as the quantum
channel; C.C. denotes the necessary classical communication cost;
I.D. denotes the entangled state needed to be identified; GHZS
denotes the GHZ state;
ES denotes the entangled state; and BS stands for the Bell state.\\
\tabcolsep 0.1 in
\begin{tabular}{ccccc} \hline
Protocol & Q.S. & S.Q.C. & C.C. & I.D. \\ \hline Shi et
al$^{[29]}$ & $\alpha|00\rangle+\beta|11\rangle$ & one GHZS & 1
bit
& 1-qubit state  \\
Liu et al$^{[30]}$ & $\alpha|00\rangle+\beta|11\rangle$ & two BSs
& 2 bits &
2-qubit ES  \\
Dai et al$^{[31]}$ & $\alpha|0000\rangle+\beta|1111\rangle$ & two
GHZSs
 & 1 bit & 2-qubit ES  \\
Zhan et al$^{[34]}$ & $\alpha|00\rangle+\beta|11\rangle$ & two BSs
& 2 bits &
2-qubit ES  \\
Wang et al$^{[33]}$ & $\alpha|000\rangle+\beta|111\rangle$ & one GHZS and one BS & 0.5 bit & 2-qubit ES \\
Our protocol &
$\alpha\prod^m_{i=1}|0\rangle_i+\beta\prod^m_{i=1}|1\rangle_i$ &
one BS & 0.5 bit & 1-qubit state \\ \hline
\end{tabular}

\vskip 0.5cm

So far one can see that all the preparation goals in Refs.[29-31,
33-34] can also be achieved by using our this RSP protocol. Now
let us make some detailed comparisons between our present protocol
and the previous protocols Refs.[29-31,33-34]. Different factors
of different protocols are summarized in Table I. The following
points can be abstracted via comparisons: \ $(p1)$  In
Refs.[29-31,33-34], the initial states taken as the quantum
channel are some complicated states, such as two Bell
states[30,34], one or more three-particle entangled states[29,31],
or their combination[33]. In contrast, the required quantum
entanglement as quantum channel in our protocol is only one Bell
state. To our knowledge, so far preparation of five-photon
entangled states has been achieved in experiment[37], however,
preparation more-photon entanglement is still desired.
Alternatively, when photon number is large, it is impossible to
prepare multi-photon GHZ states according to the present-day
technologies. Hence, in the case that $m$ is large, the method in
Ref.[29] is impossible in reality. Nevertheless, since the quantum
state (Bell state) taken as quantum channel in our protocol is
comparatively simple, the preparation and secure distribution
difficulties of the initial state taken as quantum channel is
greatly reduced. \  $(p2)$   The amount of classical communication
needed in Refs.[29-31,34] are much larger. For examples, with the
same success probability of preparation, 1 classical bit is
required by the protocols in Refs.[29,31] and two classical bits
by the protocols in Refs.[30,34]. In contrast, in this protocol,
the necessary classical communication cost is only 0.5 forward
classical bits on average. Moreover, if the state $|P\rangle$
belongs to some special states, the preparation can be realized in
a deterministically manner by consuming one extra classical bit on
average. Hence compared to the protocols [29-31,34], our protocol
consumes the minimum classical communication. $(p3)$ \ In
Refs.[30-31,33-34], Alice needs to complete the identification of
two-particle entangled states. In this protocol, the
identification Alice needs to complete is just a single-qubit
state, disregarding completely how many qubits there are. Thus, it
is obvious in our protocol the difficulty of Alice's
identification on her quantum state is degraded. According to
these comparisons, one can conclude that our protocol is much
simple, economical and feasible.

To summarize, with one Bell state as the quantum channel, we have
explicitly proposed a protocol for remotely preparing a class of
multi-qubit state by applying the CNOT gate. Although the idea in
the present scheme is simple, it can indeed realize the
preparation of the class of $m$-qubit state. More important is
that, compared to the previous protocols[29-31,33-34], our this
protocol reduces greatly the required quantum entanglement and
classical communication cost. Moreover, the difficulty of
identifying quantum state can also be degraded. Hence our
protocol is economical and feasible.\\

\noindent {\bf Acknowledgements}

This work is supported by the National Natural Science Foundation
of China under Grant Nos. 60677001 and 10304022, the
science-technology fund of Anhui province for outstanding youth
under Grant No.06042087, the general fund of the educational
committee of Anhui province under Grant No.2006KJ260B, and the key
fund of the ministry of education
of China under Grant No.206063. \\

\noindent {\bf References}

\noindent[1] C. H. Bennett, G. Brassard, C. Cr¡äepeau, R. Jozsa, A.
Peres, W. K. Wootters, Phys. Rev. Lett. {\bf70} (1993) 1895.

\noindent[2] M. Hillery, V. B\v{u}zek, A. Berthiaume, Phys. Rev. A
{\bf59} (1999) 1829.

\noindent[3] A. K. Ekert, Phys. Rev. Lett. {\bf67} (1991) 661.

\noindent[4] Z. J. Zhang {\it et al}., Eur. Phys. J. D. {\bf33}
(2005) 133; Z. J. Zhang, Phys. Lett. A {\bf342} (2005) 60.

\noindent[5] Z. J. Zhang {\it et al}., Phys. Rev. A {\bf71} (2005)
044301; Chin. Phys. Lett. {\bf22} (2005) 1588.

\noindent[6] Z. J. Zhang, Z. X. Man, Phys. Rev. A {\bf72} (2005)
022303; Phys. A{\bf361} (2006) 233; Opt. Commun. {\bf261} (2006)
199.

\noindent[7] F. G. Deng, X. H. Li, H. Y. Zhou, Z. J. Zhang, Phys.
Rev. A {\bf72} (2005) 044302. F. G. Deng {\it et al}., Phys. Rev. A
{\bf72} (2005) 044301.

\noindent[8] Z. J. Zhang et al,Phys. Lett. A {\bf 341} (2005) 55;
Commun. Theor. Phys. {\bf44} (2005) 847; Z. J. Zhang, Phys. Lett.
A {\bf352} (2006) 55.

\noindent[9] H. K. Lo, Phys. Rev. A {\bf62} (2000) 012313.

\noindent[10] A. K. Pati, Phys. Rev. A {\bf63} (2001) 014302.

\noindent[11] C. H. Bennett, D. P. DiVincenzo, P. W. Shor, J. A.
Smolin, B. M. Terhal, W. K. Wootters, Phys. Rev. Lett {\bf87}
(2001) 077902.

\noindent[12] I. Devetak, T. Berger, Phys. Rev. Lett {\bf87}
(2001) 177901.

\noindent[13] B. Zeng, P. Zhang, Phys. Rev. A {\bf65} (2002) 022316.

\noindent[14] D. W. Berry, B. C. Sanders, Phys. Rev. Lett {\bf90}
(2003) 027901.

\noindent[15] D. W. Leung, P. W. Shor, Phys. Rev. Lett {\bf90}
(2003) 127905; Z. Kurucz, P. Adam, J. Janszky, Phys. Rev. A {\bf73}
(2006) 062301.

\noindent[16] A. Hayashi, T. Hashimoto, M. Horibe, Phys. Rev. A
{\bf67} (2003) 052302.

\noindent[17] A. Abeyesinghe, P. Hayden, Phys. Rev. A {\bf68} (2003)
062319.

\noindent[18] M. Y. Ye, Y. S. Zhang, G. C. Guo, Phys. Rev. A {\bf69}
(2004) 022310.

\noindent[19] Y. F. Yu, J. Feng, M. S. Zhan, Phys. Lett. A {\bf310}
(2003) 329; Y. X. Huang, M. S. Zhan, Phys. Lett. A {\bf327} (2004)
404.

\noindent[20] M. G. A. Paris, M. Cola, R. Bonifacio, J. Opt. B
{\bf5} (2003) S360.

\noindent[21] Z. Kurucz, P. Adam, Z. Kis, J. Janszky, Phys. Rev. A
{\bf72} (2005) 052315.

\noindent[22] S. A. Babichev, B. Brezger, A. I. Lvovsky, Phys. Rev.
Lett {\bf92} (2004) 047903.

\noindent[23] G. Gour, B. C. Sanders, Phys. Rev. Lett {\bf93} (2004)
260501.

\noindent[24] D. W. Berry, Phys. Rev. A {\bf70} (2004) 062306.

\noindent[25] C. H. Bennett, P. Hayden, D. W. Leung, P. W. Shor, A.
Winter, IEEE Trans. Inform. Theory {\bf51} (2005) 56.

\noindent[26] X. Peng, X. Zhu, X. Fang, M. Feng, M. Liu, K. Gao,
Phys. Lett. A {\bf306} (2003) 271.

\noindent[27] G. Y. Xiang, J. Li, Y. Bo, G. C. Guo, Phys. Rev. A
{\bf72} (2005) 012315.

\noindent[28] N. A. Peters, J. T. Barreiro, M. E. Goggin, T. C. Wei,
P. G. Kwiat, Phys. Rev. Lett {\bf94} (2005) 150502.

\noindent[29] B. S. Shi, A. Tomita, J. Opt. B {\bf4} (2002) 380.

\noindent[30] J. M. Liu, Y. Z. Wang, Phys. Lett. A {\bf316} (2003)
159.

\noindent[31] H. Y. Dai, P. X. Chen, L. M. Liang, C. Z. Li, Phys.
Lett. A {\bf355} (2006) 285.

\noindent[32] C. S. Yu, H. S. Song, Y. H. Wang, Phys. Rev. A {\bf73}
(2006) 022340.

\noindent[33] Z. Y. Wang et al, (to be published in Commun. Theor.
Phys.)

\noindent[34] Y. B. Zhan, Phys. Lett. A {\bf336} (2005) 317.

\noindent[35] M. M. Cola, M. G. A. Paris, Phys. Lett. A {\bf337}
(2005) 10.

\noindent[36] N. B. An, Phys. Lett. A {\bf341} (2005) 9.

\noindent[37] Z. Zhao, Y. A. Chen, A. N. Zhang, T. Yang, H. J.
Briegel, J. W. Pan, Nature (London) {\bf430} (2004) 54.

\enddocument